\documentclass{aa}

\usepackage{graphicx}
\usepackage{txfonts}
\usepackage{wasysym}
\usepackage[utf8]{inputenc}
\usepackage{natbib}
\usepackage{float}
\usepackage{mathrsfs}
\usepackage{hyperref}
\usepackage{url}
\usepackage[utf8]{inputenc}
\usepackage[english]{babel}

\newcommand{\ovi}{O~{\sc{vi}}}
\newcommand{\rsun}{R$_{\odot}$}

\newcommand{\lamm}{\lambda\lambda}

\newcommand{\lya}{Ly$\alpha$}
\newcommand{\hi}{H~{\sc{i}}}
\newcommand{\kms}{km~s$^{-1}$}

\renewcommand{\vec}[1]{\mathbf{#1}}

\begin{document}

\title{Combining White Light and UV Lyman-$\alpha$ Coronagraphic Images to \\ determine the Solar Wind Speed: the Quick Inversion Method}

\author{A. Bemporad\inst{1} \and S. Giordano\inst{1} \and L. Zangrilli\inst{1} \and F. Frassati\inst{1}}

\institute{
	${^1}$ Istituto Nazionale di Astrofisica, Osservatorio Astrofisico di Torino, via Osservatorio 20, Pino Torinese 10025, Italy \\ 
	\email{alessandro.bemporad@inaf.it}}

\date{Received May 10, 2021; accepted July 2, 2021}

\abstract
{The availability of multi-channel coronagraphic images in different wavelength intervals
	acquired from the space will provide a new view of the solar corona, allowing to
	investigate the 2D distribution and time evolution of many plasma physical
	parameters, such as plasma density, temperature, and outflow speed.}
{This work focuses on the combination of White Light (WL) and UV (\lya) coronagraphic 
	images to demonstrate the capability to measure the solar wind speed in the inner corona 
	directly with the ratio between these two images (a technique called "quick inversion 
	method"), thus avoiding to account for the line-of-sight (LOS) integration effects in the
	inversion of data.}
{After a derivation of the theoretical basis and illustration of the main hypotheses in 
	the "quick inversion method", the data inversion technique is tested first with 1D
	radial analytic profiles, and then with 3D numerical MHD simulations, in order to show
	the effects of variabilities related with different phases of solar activity cycle and
	complex LOS distribution of plasma parameters. The same technique is also applied to 
	average WL and UV images obtained from real data acquired by SOHO UVCS and LASCO
	instruments around the minimum and maximum of the solar activity cycle.}
{Comparisons between input and output velocities show overall a good agreement,
	demonstrating that this method that allows to infer the solar wind speed with WL-UV 
	image ratio can be complementary to more complex techniques requiring the full LOS integration. 
	The analysis described here also allowed us to quantify the possible errors in the outflow speed, 
	and to identify the coronal regions where the "quick inversion method" performs at the best. The
	"quick inversion" applied to real UVCS and LASCO data allowed also to reconstruct the typical bimodal 
	distribution of fast and slow wind at solar minimum, and to derive a more complex picture around 
	solar maximum.}
{The application of the technique shown here will be very important for the future analyses 
	of data acquired with multi-channel WL and UV (\lya) coronagraphs, such as Metis on-board
	Solar Orbiter, LST on-board ASO-S, and any other future WL and UV \lya\ multi-channel 
	coronagraph.}

\keywords{Sun: corona -- techniques: polarimetric -- methods: data analysis -- solar wind -- Sun: UV radiation}

\titlerunning{Solar Wind Velocity Measurements with the Quick Inversion Method}

\authorrunning{Bemporad et al.}

\maketitle

\section{Introduction} \label{sec: introduction}

Near the Sun, where the main acceleration of solar wind from sub- to super-sonic and super-Alfv\'enic flows occur \citep[below $\sim 15-30$ \rsun, see e.g.][]{goelzer2014}, measurements of the expansion speed of solar wind have been possible so far only with remote sensing data. The obvious reason for this limitation is that, due to the extreme local conditions, there were no instruments capable to explore \textit{in situ} this region. Most recently, thanks to the launch of the Parker Solar Probe mission \citep{fox2016}, it became possible to explore for the first time with \textit{in situ} instruments regions much closer to the Sun (down to $\sim 10$ \rsun), but the exploration of the inner regions still requires the analysis of remote sensing data. Currently available coronagraphic data have already proven their potential, but also their limits. In particular, classical space-based coronagraph, such as the instrument on-board the Solar Maximum Mission \citep{macqueen1980}, LASCO on-board SOHO \citep{brueckner1995}, COR on-board STEREO \citep{howard2008}, were limited so far to the acquisition of broad-band images in the White Light (WL). This emission, being mostly due to Thomson scattering of photospheric light from coronal electrons, is very useful to observe large-scale coronal features, but provides local information only on the plasma column density along the line-of-sight (LOS), and no local information on other plasma parameters (temperatures of different plasma species, elemental abundances, etc.).

Fortunately, this situation will change in the near future with the new generation of multi-waveband coronagraphs, whose data will provide a new view of the solar corona, and in particular of the inner regions where the main solar wind acceleration and coronal heating processes occur. In particular, the Metis coronagraph \citep{antonucci2020} on-board ESA Solar Orbiter mission is now providing the first ever simultaneous observations of the corona in two different spectral bands: broad-band (580-640 nm) in the WL, and narrow-band UV emission from the neutral H atoms (121.6 nm \lya\ line), with a FOV going from 1.7 to 3.6 \rsun\ at closest approach (0.28 AU) to the Sun. Similar data will also be acquired in the near future by another coronagraph, the LST instrument \citep{li2019} on-board the forthcoming Chinese ASO-S mission. The great advantages for solar wind diagnostic in the combination of UV \lya\ and WL $pB$ coronal images was first discussed by \citet{withbroe1982} who proposed to measure the expansion speed of the neutral H atoms with the so-called Doppler dimming technique. The measurement is based on well known fact that the coronal \lya\ emission is almost entirely due to the resonant scattering of chromospheric \lya\ emission \citep{gabriel1971}. As a consequence, a relative motion of coronal H atoms with respect to the chromosphere leads to a Doppler shift of the exciting \lya\ profile with respect to the atomic absorption profile, reducing the efficiency of radiative excitation, and then the observed \lya\ emission \citep[see][for a more recent review]{vial2016}. This method is suitable for coronal features such as streamers, plumes, and coronal holes, where (thanks to the low plasma densities) the collisional \lya\ component is negligible, which is not necessarily true for the emission associated with erupting prominences and the cores of Coronal Mass Ejections where the \lya\ collisional component can be significant or event 
dominant, as demonstrated by more recent data analysis \citep{susino2018} and numerical MHD simulations \citep{bemporad2018}.

The Doppler dimming technique has been successfully applied to measure the solar wind speed in the inner corona with the analysis of spectro-coronagraphic observations acquired by the UVCS instrument \citep{kohl1995} on-board SOHO \citep[see][for a review of the main UVCS results]{kohl2006}. This kind of measurements also increases the complementarity between \textit{in situ} and remote sensing data, allowing to relate plasma physical parameters measured locally with their global large-scale distribution. Nevertheless, the application of the Doppler dimming technique usually requires not only the integrated intensity of some selected spectral lines (in particular the 121.6 nm \hi\ \lya\ line, and the 103.2--103.8 nm \ovi\ doublet lines), but also information on the shape of coronal profiles. Unfortunately, this information will not be provided by a coronagraph like Metis, which is not equipped with a spectroscopic channel like UVCS. Nevertheless, reliable solar wind speed measurements can be derived with Doppler dimming technique even just from the \lya\ integrated intensity, once a corresponding WL coronagraphic image is provided. This was more recently demonstrated in \citet{bemporad2017}, who derived 2D maps of the solar wind speed using the so-called synoptic UVCS observations \citep{strachan1997, panasyuk1998} to build 2D coronagraphic \lya\ intensity images, and then coupling these images with WL coronagraphic observations. The same images were thus used by \citet{dolei2018} and \citet{dolei2019} to characterize and constrain the main uncertainties in the Doppler dimming technique applied to 2D intensity images, missing the spectroscopic information.

Nevertheless, the methods applied first by \citet{bemporad2017}, and then by \citet{dolei2018} and \citet{dolei2019} to the same reconstructed images were intrinsically different, as it is explained in more details below. In summary, the method applied by \citet{bemporad2017} (hereafter called "quick inversion method") was based on the analysis of the image resulting from the direct ratio between the UV and WL images, while the method applied by \citet{dolei2018} and \citet{dolei2019} (hereafter called "full inversion method") was based on a more complete analysis taking into account LOS integration effects. The aim of this paper is to review the two methods, and in particular to derive explicit expressions useful to apply the "quick inversion method" to future observations of the corona both in the WL and UV \lya\ bands. In particular, after a short review of the two methods (\S~\ref{sec: methods}), the approximate expressions to be used are discussed in details (\S~\ref{sec: approximate}) both for the UV \lya\ (\S~\ref{sec: lya}) and the WL $pB$ (\S~\ref{sec: pB}) images, deriving an explicit expression to measure the outflow speed and its associated uncertainty (\S~\ref{sec: V0}). The method is finally tested with 1D synthetic emission profiles (\S~\ref{sec: 1Dtest}), 2D synthetic images based on 3D MHD simulations (\S~\ref{sec: 3Dtest}), and 2D images based on real observations (\S~\ref{sec: realdata}). The results, advantages and limits of the method are finally summarized and discussed (\S~\ref{sec: summary}).

\section{The full and quick inversion methods} \label{sec: methods}

\subsection{Full inversion} \label{sec: full}

The Doppler dimming/pumping technique has been applied in the past mostly to measure the expansion speed 
of heavy ions, and in particular of O$^{5+}$ ions in coronal holes and coronal streamers
\citep[see reviews by][and references therein]{cranmer2002, abbo2016}, by exploiting the fact that the 
ratio between \ovi\ $\lamm$ 1031.9-1037.6 lines is mostly dependent on the ion outflow speed \citep{noci1987}, 
that can be measured once the plasma electron temperature $T_e$ and density $n_e$ (plus other parameters) are known
\citep[see][for a detailed description of the method]{zangrilli2002}. Nevertheless, the expansion speed of
heavy ions $V_{ion}$ is not representative of the solar wind bulk speed $V_{sw}$, because these ions undergo a preferential
heating and acceleration, that was one of the major discoveries by UVCS experiment \citep[see review by][]{kohl2006}.
On the other hand, the speed of the expanding protons $V_p$ can be measured almost directly by applying the
Doppler dimming technique to the \hi\ \lya\ line intensity, under the assumption that protons and neutral H atoms
are still coupled (hence $V_p \simeq V_H$), which is true only in the inner corona typically below $\sim 10$ \rsun\ \citep{withbroe1982}.
A decoupling in the direction perpendicular to the flow may occurr even at lower altitudes
\citep[see discussion by][]{allen1998}, thus possibly affecting the reliability of proton kinetic 
temperature measurements form \lya\ line profile \citep[e.g.][]{labrosse2006}.

Because coronal plasma are usually optically thin \citep[see review by][]{bradshow2013}, the derivation 
of proton speed $V_p$ on the plane-of-sky (POS) with the Doppler dimming of \lya\ intensity usually requires a 
very complex inversion procedure, taking into account not only the (unknown) distributions along the line-of-sight 
(LOS) of $T_e$, $n_e$, $T_{p//}$ and $T_{p\perp}$ (i.e. temperatures parallel and perpendicular to the magnetic 
fieldlines), but also the LOS distribution of $V_p$ itself. This problem is usually solved by estimating the POS
values of $T_e$, $n_e$ and $T_{p\perp}$, by assuming a spherical geometry for the LOS distributions of these
parameters, by starting from a level of temperature anisotropy, and also by assuming the mass flux conservation in 
magnetic fluxtubes, hence an analytical \citep[e.g.][]{banaszkiewicz1998} or an empirical \citep[e.g.][]{vasquez2003} 
model for the LOS distribution of magnetic fieldlines.
An iteration over the space of the free parameters (e.g. the POS speed and the anisotropy level) is then performed, 
until the best match between the synthetic and the observed \lya\ intensities is obtained \citep[see][for details]{zangrilli2002}.
A slightly simplified version of this technique was applied by \citet{spadaro2007} and more recently by \citet{dolei2018} 
and \citet{dolei2019}, where (considering that in particular in coronal streamers the super-radial expansion of solar 
wind hence of magnetic fluxtubes is not very important), the assumption of a LOS distribution of magnetic fluxtubes 
was replaced by the assumption that the radial POS distribution of outflow speed equals the distribution along the LOS (which 
corresponds to assume spherical symmetry also for the velocity distribution along the LOS).

\subsection{Quick inversion} \label{sec: quick}

The above methods have been applied by many authors to measure $V_p$ from the Doppler dimming of \lya\ intensity 
\citep[see Introduction and Figure 1 by][for a quick review]{bemporad2017}. Nevertheless, the significant number of 
needed assumptions on the LOS distributions of plasma parameters, results in significant uncertainties in the derived 
values for $V_p$. Recently, \citet{cranmer2020} re-analysed a large set of \lya\ line profile measurements acquired 
by UVCS in 1996-1997 in a polar coronal hole to constrain (with forward modeling based on Monte Carlo method to build a posterior 
probability distributions) the set of plasma parameters that give the best match with the UVCS data. Surprisingly, this complete 
re-analysis of data previously analysed by \citet{cranmer1999} shows that between 1.5 and 4 \rsun\ the electron temperature $T_e$, 
the proton temperature $T_p$, and the temperature anisotropy $T_{p\perp}/T_{p//}$ do not show substantial radial dependences, 
changing only by $\sim 20\%$, $\sim 84\%$, and $\sim 25\%$ respectively, hence always less than a factor $\sim 2$. The other plasma parameters 
are changing much more significantly over the same altitude interval: the proton outflow speed $V_p$ increases by a factor $\sim 4$, and more 
importantly the electron density $n_e$ decreases by almost two orders of magnitude. Moreover, coronal protons have a very little temperature
anisotropy, on the order of $T_{p\perp}/T_{p//} \sim 1.06$ \citep[see][]{cranmer2020}. The same conclusions are expected
to be valid also in coronal streamers, from where the slow solar wind is blowing \citep[see][and references therein]{kohl2006},
and where the level of proton anisotropy is expected to be negligible everywhere except in the streamer edges and coronal 
	hole boundaries \citep{frazin2003, susino2008}.

The above lines of reasoning show that, considering the LOS contribution to the coronal \lya\
intensity, there is only one plasma physical parameter dramatically changing (i.e. by orders of magnitude)
with the radial distance from the Sun: the plasma density. For this reason, over the past decades some authors used 
approximate expressions for the \lya\ intensity that are considering only the LOS variations of $n_e$, and by assuming 
that no significant LOS variations of all other plasma parameters occur. Similar approximate expressions have been
used for instance to measure the electron density at the base of coronal streamers \citep[][]{ko2002} by
assuming negligible outflows (hence neglecting Doppler dimming), but also to measure the electron temperature 
in Coronal Mass Ejections - CMEs \citep{susino2016, ying2020}, by measuring their expansion speed from coronagraphic images 
to constrain the \lya\ Doppler dimming. 

An approximate expression for the \lya\ line can be derived starting from the important 
consideration that in the solar corona the \lya\ emission is dominated by resonant scattering of chromospheric \lya\ emission
\citep{gabriel1971}. Hence, as it was originally proposed by \citet{kohl1982},
the coronal \lya\ line can be used to measure the solar wind outflow speed directly from the ratio between 
its intensity and the intensity of the electron-scattered WL continuum. For this purpose, it is necessary 
to approximate the \lya\ resonant scattering component $I_{res}$ as
\begin{equation}
	I_{res} = \mathrm{const}\,\, \langle R(T_e) \rangle \, \langle D(V_{sw}) \rangle \int\limits_{LOS}n_e\, dz \label{eq:approx1}
\end{equation}
where $R(T_e)$ is the H ionization fraction (dependent on the electron temperature $T_e$), 
$V_{sw}$ is the radial component of the wind speed, $D$ is the Doppler dimming term ($0 \leq D \leq 1$, with $D = 1$ 
for $V_{sw} = 0$, and $D \longrightarrow 0$ for $V_{sw} \longrightarrow \infty$), and the symbol $\langle ... \rangle$ 
indicates the average value along the LOS across the emitting plasma; an explicit form of $D$ will be derived below.

The intensity of the polarized brightness of WL $pB$ can also be approximated as
\begin{equation}
	I_{pB} = \mathrm{const} \int\limits_{LOS}n_e\, dz,
\end{equation}
therefore, the ratio between the two measured quantities is almost independent on $n_e$, and corresponds to
\begin{equation}
	\frac{I_{res}}{I_{pB}} = \mathrm{const}\,\, \langle R(T_e) \rangle \, \langle D(V_{sw}) \rangle.
\end{equation}
The above expression implies that, given an estimate of $T_e$, the $I_{res}/I_{pB}$ intensity ratio can be used
to measure directly $D(V_{sw})$, hence to estimate $V_{sw}$. The same technique was more recently 
described again by \citet{kohl2006} and successfully applied by \citet{bemporad2017}.

This method is relatively simple, because does not require performing iterations along the LOS based on 
geometrical assumptions of the unknown LOS distributions of many plasma physical parameters, as
previously discussed; for this reason, it is called "quick inversion method". In the next section 
we derive the explicit expression for the \lya\ approximate intensity used by this method.

\section{Approximate expressions} \label{sec: approximate}

\subsection{Approximate \lya\ expression} \label{sec: lya}

A complete review of EUV line formation processes in the solar corona was recently given by \citet{delzanna2018}.
Here, we derive the \lya\ intensity approximate form starting from the expression \citep[given by][Eq. 9]{nocimaccari1999}
for the line emissivity $j(P,\vec{n})$ of a radiatively excited line emitted from a scattering atom located at the 
coronal point $P$, after integration over the (Maxwellian) velocity distribution, which is given by
\begin{equation}
	j(P,\vec{n}) = h\nu_0\,n_H \frac{B_{12}}{4 \pi} \!\int\limits_\Omega \!\! p(\phi) d\omega' \!\!\! \int\limits_{-\infty}^{+\infty} 
	\!\! I(\nu' [V_p,\nu_0],\vec{n'}) g_p(V_p) dV_p \label{eq:convol}
\end{equation}
where $n_H$ is the number density of the absorbing H atoms, $B_{12}$ is the Einstein coefficient relative to the considered transition at 
frequency $\nu_0$, $\nu' = \nu_0 (1+\vec{n'}\cdot\vec{v}/c)$ is the frequency of the absorbed photon for an observer at rest with the 
scattering atom moving with velocity $\vec{v}$ with respect to the emitting source (the Sun), $p(\phi)$ is a geometrical factor 
representing the probability for the photon coming from the direction $\vec{n'}$
to be absorbed and re-emitted through the angle $\phi$, $\vec{n}$ is the LOS direction, $\Omega$ is the solid 
angle subtended by the solar disk at scattering point $P$, and $V_p = \vec{v}\cdot\vec{n'}$ is the velocity
of scattering atom projected in the direction of the incoming photon $\vec{n'}$ and having normalized Maxwellian distribution 
$g_p$. The line emissivity at point $P$ needs to be integrated along the LOS coordinate $z$ to get the observed resonant scattered
intensity $I_{res}$ which is given by
\begin{equation}
	I_{res} = \int\limits_{-\infty}^{+\infty} j(P,\vec{n}) dz
\end{equation}
The above line emissivity $j$ can be integrated analytically by making a few assumptions described below.
First, we assume that the excitation profile has a Gaussian shape, and thus is given by \citep[see][Eq. 10]{nocimaccari1999}
\begin{equation}
	I(\nu' [V_p,\nu_0],\vec{n'}) = \frac{I_0}{\sqrt{\pi}\sigma_\nu} \, \exp\left[ -\frac{(\nu'-\nu_0)^2}{\sigma_\nu^2}  \right]
\end{equation}
and considering that $\nu' = \nu_0 (1-V_p/c)$ the above profile can be rewritten as
\begin{eqnarray}
	I(V_p) 	&=& \frac{I_0\, \lambda_0}{\sqrt{\pi}(c\,\sigma_\nu /\nu_0)} \, \exp\left[ -\frac{V_p^2}{(c\,\sigma_\nu /\nu_0)^2}  \right]= \\ \nonumber
			&=& \frac{I_0\, \lambda_0}{\sqrt{\pi}s_{disk}} \, \exp\left[ -\frac{V_p^2}{s_{disk}^2}  \right] \label{eq:excit}
\end{eqnarray}
where $s_{disk} = \sigma_\nu \,c/\nu_0$ [\kms] is $1/e$ half-width of the disk exciting profile with total intensity $I_0$. 
On the other hand, by assuming that the scattering atoms are moving radially with the bulk solar wind velocity $V_0$ 
\citep[using the same notation as][]{nocimaccari1999}, the normalized atomic absorption profile $g_p(V_p)$ can be written as
\begin{equation}
	g_p(V_p) = \frac{1}{\sqrt{\pi}s_{cor}} \, \exp\left[ -\frac{(V_p-V_0)^2}{s_{cor}^2}\right] \label{eq:absprof}
\end{equation}
where $s_{cor} = \sqrt{2\,k_B T_k/m_H + \xi^2}$ [\kms] is the $1/e$ half-width of the absorption profile, that includes the
thermal ($2\,k_B T_k/m_H$) and non-thermal ($\xi^2$) line broadening, $T_k$ is the kinetic temperature of the scattering ions. 
Note that Eq. \ref{eq:absprof} is valid in the assumption of an isotropic temperature distribution. By replacing 
the above expressions for $I(V_p)$ and $g_p(V_p)$ the convolution integral in Eq. \ref{eq:convol} becomes:
\begin{equation}
	I_0\lambda_0 \int\limits_{-\infty}^{+\infty} \frac{1}{\sqrt{\pi}s_{disk}}e^{-V_p^2/s_{disk}^2} \frac{1}{\sqrt{\pi}s_{cor}}
	e^{-(V_p-V_0)^2/s_{cor}^2}\,dV_p. \label{eq:convol2}
\end{equation}

The above expression can be integrated analytically: as it is well-known \citep[e.g.][]{bromiley2014} the product between two normalized 
Gaussian functions $g(V)$ with peak values at $V_1$ and $V_2$ and $1/e$ half-widths $s_1$ and $s_2$ is a normalized Gaussian function with 
mean $V_{12} = (V_1s_2^2+V_2s_1^2)/(s_1^2+s_2^2)$, variance $s_{12} = \sqrt{s_1^2s_2^2/(s_1^2+s_2^2)}$, and multiplied 
by the constant
\begin{equation}
	G_{12} = \frac{1}{\sqrt{\pi}\sqrt{s_1^2 + s_2^2}} \exp{\left[-\frac{(V_1-V_2)^2}{(s_1^2+s_2^2)} \right]}
\end{equation}
where in our case $V_1=0$ and $V_2=V_0$. Hence, the integration of Eq. \ref{eq:convol2} gives
\begin{eqnarray}
	\int\limits_{-\infty}^{\infty} \!\! I(\nu' [V_p,\nu_0],\vec{n'}) g_p(V_p) dV_p =  
	\frac{I_0\lambda_0\,\exp\left[ -\frac{V_0^2}{(s_{disk}^2+s_{cor}^2)}\right]}{\sqrt{\pi}\sqrt{s_{disk}^2 + s_{cor}^2}} . \label{eq:convol3}
\end{eqnarray}

The expression given in Eq. \ref{eq:convol} can be further simplified by considering that for the \lya\ line \citep{noci1987}
\begin{equation}
	p(\phi) = \frac{1}{4\pi}\frac{11+3 \cos^2\phi}{12} \simeq \frac{1}{4\pi}
\end{equation}
because the bulk of the \lya\ emission is coming from the plasma located near the POS ($\cos \phi \simeq 0$) and 
the integration over the solid angle $d\omega'$ can be factorized and simplified into
\begin{equation}
	\int\limits_\Omega \!\! p(\phi) d\omega' \simeq \frac{1}{4\pi} \Omega_\odot(\rho) = \frac{1}{4} h(\rho) \label{eq:omega}
\end{equation}
where $\Omega_\odot(\rho)$ is the solid angle subtended by the solar disk at scattering distance $\rho$
on the POS, and the function $h(\rho)$ is given by $h(\rho) = 2\left( 1-\sqrt{1-R_\odot^2/\rho^2}\right)$ \citep[see e.g.][]{ko2002}.
Moreover, the number density of neutral H atoms can be rewritten as
\begin{equation}
	n_H \equiv \frac{n_H}{n_p}\frac{n_p}{n_e}\,n_e = R_H(T_e) \frac{n_p}{n_e}\,n_e \simeq R_H(T_e)\,0.83\,n_e \label{eq:ioniz}
\end{equation}
where $R_H(T_e)$ is the neutral Hydrogen ionization fraction, and $n_p/n_e = n_H/(n_H+2n_{He}) \equiv 1/(1+2n_{He}/n_H) \simeq 0.83$ 
having assumed \citep[e.g.][]{delzanna2018} that the considered plasma is made of a combination of 90\% of Hydrogen and 10\% 
of Helium ($n_{He}/n_H \simeq 0.1$).

\begin{figure}[t!]
	\centering
	\includegraphics[width=0.5\textwidth]{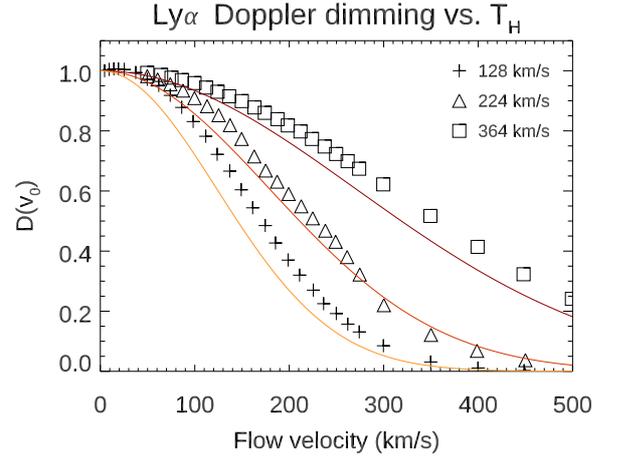}
	\caption{
		Comparison between the \lya\ Doppler dimming factors as provided by \citet{kohl1997} (Fig. 1, same symbols)
		and those obtained with the analytic formula given here in Eq. \ref{eq: explicitdd} (solid lines).
	}
	\label{Fig1}
\end{figure}
In summary, by replacing Eqs. \ref{eq:convol3}, \ref{eq:omega}, and \ref{eq:ioniz} into Eq. \ref{eq:convol} it turns out that
\begin{eqnarray}
	j(\rho) = 0.83\frac{h\,\nu_0 B_{12}}{16\pi\sqrt{\pi}}R_H\left[T_e(\rho)\right] n_e(\rho)h(\rho) \times \\ \nonumber
	\times \frac{I_0\lambda_0}{\sqrt{s_{disk}^2+s_{cor}^2}}
	\exp\left[ -\frac{V_0^2}{(s_{disk}^2+s_{cor}^2)}\right]
\end{eqnarray}
that can be also rewritten by replacing the observed $1/e$ half-widths of the excitation ($\sigma_{disk}(\lambda)$) and 
absorption ($\sigma_{cor}(\lambda)$) profiles (both assumed to be Gaussian) as
\begin{eqnarray}
	j(\rho) = 0.83\frac{h\,\lambda_0 B_{12}}{16\pi\sqrt{\pi}}R_H\left[T_e(\rho)\right] n_e(\rho)h(\rho) \times \\ \nonumber
	\times \frac{I_0}{\sqrt{\sigma_{disk}^2+\sigma_{cor}^2}} \exp\left[ -\frac{V_0^2}{(\sigma_{disk}^2+\sigma_{cor}^2)c^2/\lambda_0^2}\right]
\end{eqnarray} 
and finally, the above expression needs to be integrated along the LOS coordinate $z=\sqrt{r^2-\rho^2}$ to get the total intensity. 

Moreover, the above equation provides a useful analytic expression for the Doppler dimming coefficient $D(V_0)$ which is given by
\begin{equation}\label{eq: explicitdd}
	D(V_0) = \exp\left[ -\frac{V_0^2}{(\sigma_{disk}^2+\sigma_{cor}^2)c^2/\lambda_0^2}\right].
\end{equation}
This is a general expression (under all the above assumptions) that holds for any coronal spectral line radiatively 
excited by the disk emission in the same line. A comparison
between values for $D(V_0)$ given by the above analytic expression (by assuming $1/e$ half-width of the \lya\ chromospheric 
profile $\sigma_{disk} = 0.34$ \AA) and those given by \citet{kohl1997} is provided in
Fig. \ref{Fig1}, showing a nice agreement between the different curves for different kinetic temperatures $T_k$ 
within $\sim 20-30$ \kms. This functional form of Doppler dimming coefficient was the same already used for instance 
by \citet{cranmer1999} (Eq. 17) to measure the solar wind speed in polar coronal holes. Also note
that, because both the excitation and absorption profiles were assumed to be Gaussian hence symmetric, the above expression for 
the Doppler dimming coefficient holds both for plasma escaping from the Sun ($V_0 = V_{sw} > 0$), and for plasma moving towards
the Sun, as it may happen to down-flowing plasma blobs \citep[e.g.][]{wang1999}, but also for the cometary emission by 
considering the radial component ($V_0 = V_{r;com} \gtrless 0$) of the comet velocity, an effect called "Swings effect" \citep{swings1941}.

\begin{figure}[t!]
	\centering
	\includegraphics[width=0.5\textwidth]{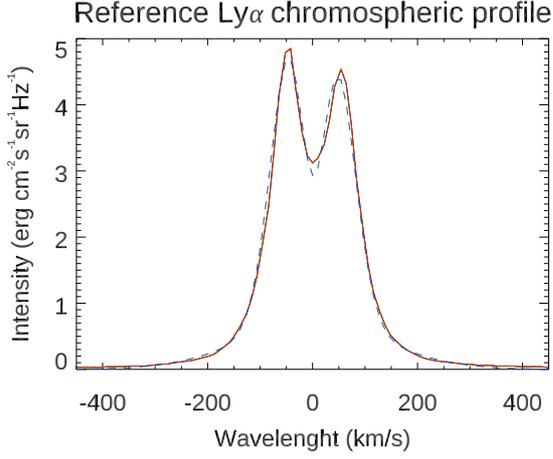}
	\caption{
		Reference \lya\ chromospheric profile from \citet{gunar2020} (solid line)
		and three-Gaussian fitting (dashed line).
	}
	\label{Fig2}
\end{figure}
As discussed above,
in the approximate expression considered here, it will be assumed that all the main plasma physical parameters are not
changing significantly along the LOS in the emitting plasma column, with the only exception to the electron density $n_e$,
so that the total scattered intensity $I_{res}(\rho)$ at the projected distance $\rho$ from the Sun is given by
\begin{eqnarray}\label{eq: uvbestapprox}
	I_{res}(\rho) = 0.83\frac{h\,\lambda_0 B_{12}I_0}{16\pi\sqrt{\pi}}\, \frac{R_H\left[T_e(\rho)\right]\,h(\rho) }{\sqrt{\sigma_{disk}^2+\sigma^2_{cor}(\rho)}} 
	D[V_0(\rho)]\times \\ \nonumber 
	\times \int\limits_{-\infty}^{+\infty} n_e(z) dz = H_{res}\,K_{res}(\rho)\,D[V_0(\rho)]\,\int\limits_{-\infty}^{+\infty}\,n_e(z)\,dz
\end{eqnarray}
where in the above expression all the varying quantities are written as a function of $\rho$. 

It is important to notice that, as any approximation, the above expression has some limits. In particular, 
	the assumption that the integration over the solid angle subtended by the solar disk can be simply factorized as 
	expressed by Eq. \ref{eq:omega} (usually referred as "point source" approximation) fails for regions in the inner 
	corona where different values of this solid angle in the integration along the LOS need to be taken into account.
	The assumption that the same value of solid angle $\Omega_\odot(\rho)$ applies also for plasma emitting out from 
	the POS leads to an overestimate of the emission from these coronal regions. A correction for the errors introduced 
	by this approximation will be discussed later.

\begin{figure*}[t!]
	\centering
	\includegraphics[width=0.75\textwidth]{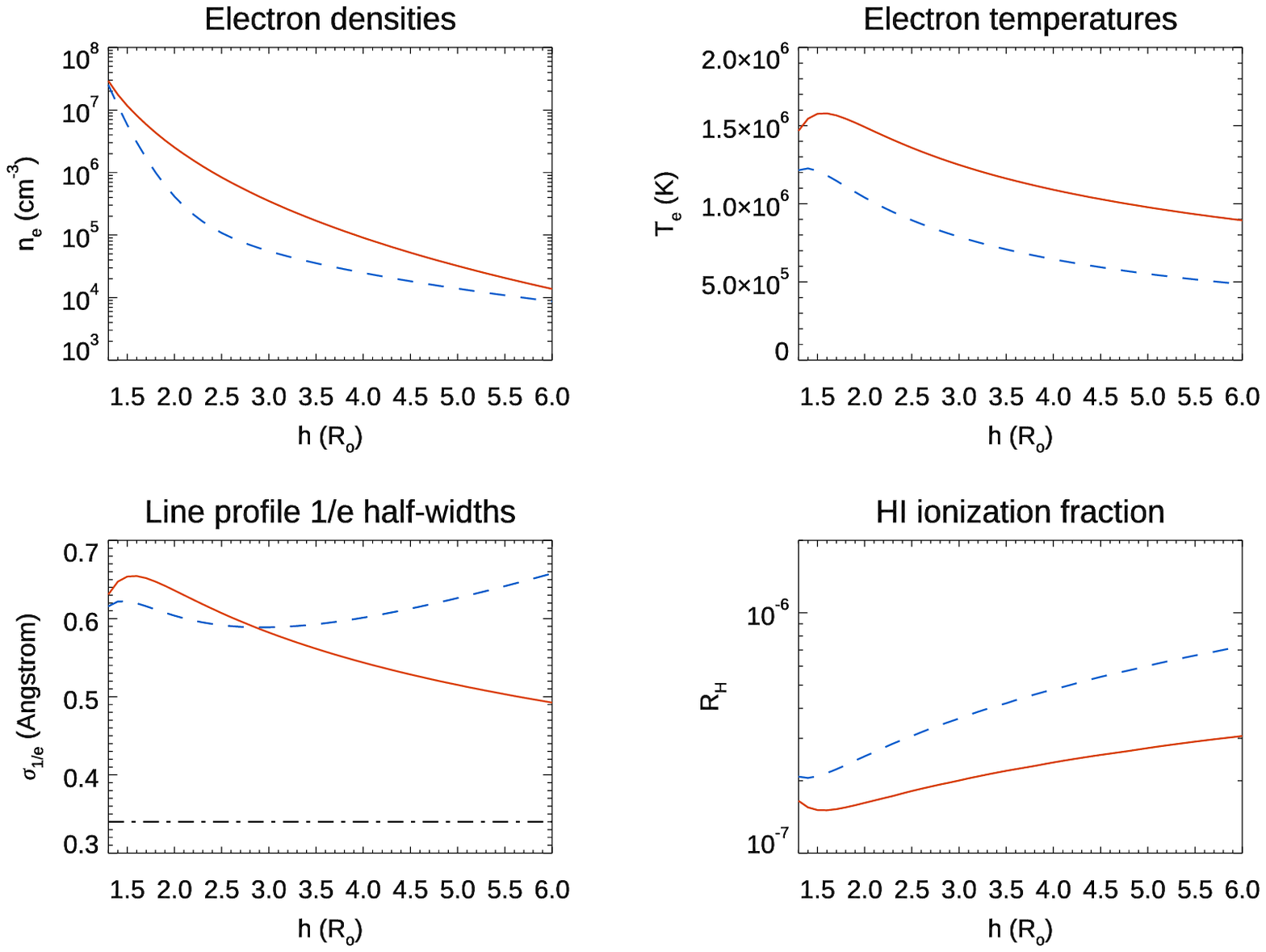}
	\includegraphics[width=0.75\textwidth]{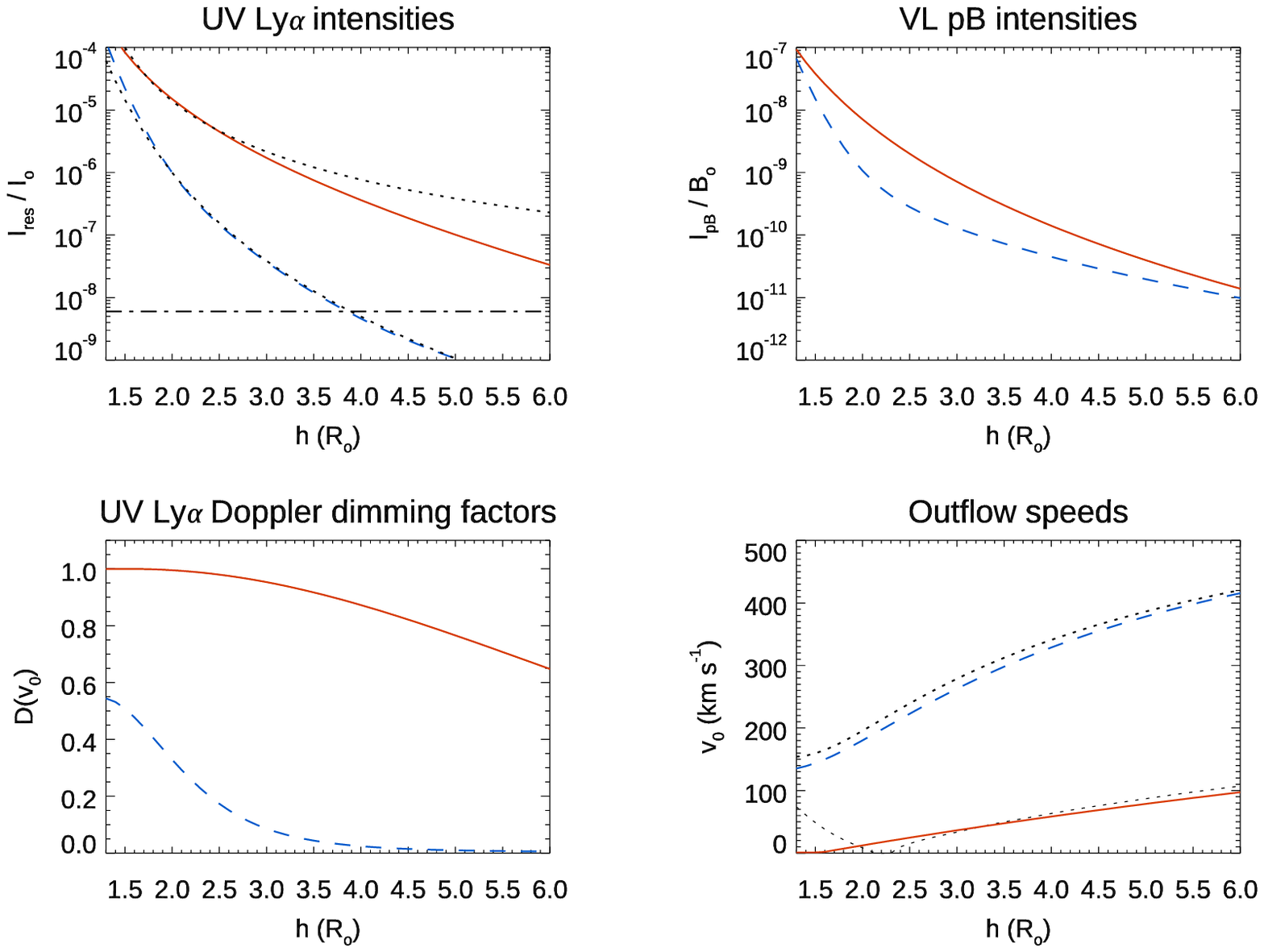}
	\caption{
		Radial profiles of different input and output parameters in coronal streamer (solid red lines) and coronal
		hole (dashed blue lines), in particular (from top left to bottom right): electron densities, electron
		temperatures, \lya\ line profile widths (the dash-dotted line shows the chromospheric line widths), \hi\
		ionization fractions, $I_{res}$ \lya\ intensities (normalized with respect to the disk intensity $I_0$, 
		the dash-dotted line shows the interplanetary \lya\ intensity, while dotted lines show typical \lya\ intensity
		profiles as observed by SOHO UVCS at solar minimum), $I_{pB}$ intensities (normalized with
		respect to the mean solar brightness $\bar{B}_{\odot}$), $I_{res}$ \lya\ Doppler dimming factors, and
		input and output (dotted lines) outflow speed ($V_0$) profiles.
	}
	\label{Fig3}
\end{figure*}
In the analysis described above it was also assumed (as done by many previous authors) that the \lya\ chromospheric profile
can be well approximated by a single Gaussian profile. Nevertheless, as it is well-known, the \lya\ chromospheric emission
is characterized by a reversed shape around the line centre due to the absorption from H in the upper chromosphere and 
transition region; a reference \lya\ disk profile was recently provided by \citet{gunar2020}. As it was shown for
instance by \citet{auchere2005}, this profile can be well approximated by the superposition of three Gaussian profiles,
that can be replaced in the expression for the excitation profile (Eq. \ref{eq:excit}), hence in the convolution 
integral (Eq. \ref{eq:convol2}). More recently, a reference full-disk \lya\ chromospheric profile was published by 
\citet{gunar2020}; starting from the profile provided by these authors, the single Gaussian profile can be replaced by
\begin{eqnarray}
	I(V_p) 	= \frac{\lambda_0}{\sqrt{\pi}} \,\left[ i_0 e^{-\frac{V_p^2}{s_{disk0}^2}} + 
	i_1 e^{-\frac{V_p^2}{s_{disk1}^2}} + i_2  e^{-\frac{(V_p-V_2)^2}{s_{disk2}^2}} \right] \label{eq:excit2}
\end{eqnarray}
having assumed that only one of the three Gaussian profiles is shifted with respect to the reference central wavelength 
$\lambda_0$. In particular, the fitting curve (shown in Fig. \ref{Fig2}) corresponds to $i_0 = 5.712$, $i_1 = 0.930$,
and $i_2 = -3.085$ erg cm$^{-2}$s$^{-1}$sr$^{-1}$Hz$^{-1}$, $s_{disk0} = 68.67$, $s_{disk1} = 193.52$,  $s_{disk2} = 46.11$
\kms, and $V_2 = 0.914$ \kms. By replacing the above expression for the exciting profile $I(V_p)$ into the convolution 
integral (Eq. \ref{eq:convol2}) it is possible to derive a more refined expression for the Doppler dimming coefficient
$D[V_0(\rho)]$, that will result in the sum of three exponential terms similar to the single one given in Eq. \ref{eq: explicitdd},
whith the only disadvantage that it is not possible any more to derive an explicit solution for the plasma flow velocity $V_0$.


\subsection{Approximate pB expression} \label{sec: pB}

The derivation of an approximate expression for the WL polarized brightness ($pB$) is more straightforward, because 
this quantity depends only on the integration along the LOS of $n_e$ multiplied by some geometrical functions depending only
on the heliocentric distance $r$. In particular, by assuming that in the available $pB$ image the F-corona emission due to 
scattering by interplanetary dust has been entirely removed, hence the K-corona emission has been isolated, the $I_{pB}(\rho)$ 
intensity observed at the projected distance $\rho$ on the POS is given by \citep{vandehulst1950}
\begin{equation} \label{eq:pb definition}
	I_{pB}(\rho) = \frac{\pi}{2}\,\sigma_T\,\bar{B}_{\odot}\,
	\int\limits_{-\infty}^{+\infty}\,n_e(z)\,\left[ \frac{(1-u)A(r)+uB(r)}{1-u/3} \right] 
	\,\frac{\rho^2}{r^2}\,dz
\end{equation}
where $z=\sqrt{r^2-\rho^2}$, $u=0.63$ is the limb darkening coefficient in the visible wavelength
of interest, $\bar{B}_{\odot}$ is the mean solar brightness in the considered wavelength band, and 
the expressions for functions $A(r)$ and $B(r)$ are given by \citet{billings1966}.
In order to apply the quick inversion method described in the previous section, it is necessary to assume that the 
above expression can be simplified into
\begin{equation}
	I_{pB}(\rho) = \frac{\pi}{2}\,\sigma_T\,\bar{B}_{\odot}\,K_{pB}(\rho)
	\int\limits_{-\infty}^{+\infty}\,n_e(z)\,dz
\end{equation}
where $K_{pB}(\rho) = [(1-u)A(\rho)+ uB(\rho)]/(1-u/3)$. Defining the constant 
quantity $H_{pB} = \pi/2\, \sigma_T\,\bar{B}_{\odot}$, the polarized brightness will be 
conveniently approximated by
\begin{equation}\label{eq: wlbestapprox}
	I_{pB}(\rho) = H_{pB}\,K_{pB}(\rho) \int\limits_{-\infty}^{+\infty}\,n_e(z)\,dz.
\end{equation}
an expression that will be used in the next Section to derive an explicit expression
for the outflow speed $V_0$.

Similar to \lya, also the above approximate expression for $pB$ may lead to wrong estimates of the expected intensity.
	In particular, as it is possible to verify numerically by assuming well-established electron density radial profiles $n_e(r)$ 
	from the literature for coronal streamers \citep{gibson1999} and coronal holes \citep{cranmer1999}, this approximated expression
	for $pB$ provides in general an overestimate by a factor of $1.2-1.6$. This may lead to wrong estimates of the outflow
	velocity $V_0$; corrections for these errors will be discussed later.

\subsection{Outflow velocity measurement} \label{sec: V0}

Finally, Eqs. \ref{eq: uvbestapprox} and \ref{eq: wlbestapprox} can be combined into
\begin{equation}
	\frac{I_{res}(\rho)}{I_{pB}(\rho)} = \frac{H_{res}}{H_{pB}}\frac{K_{res}(\rho)}{K_{pB}(\rho)}D[V_0(\rho)]
	\label{eq: DDratios}
\end{equation}
and, by using Eq. \ref{eq: explicitdd}, it is possible to derive an explicit form for the velocity $V_0$ given by
\begin{equation}
	V_0(\rho) = \sqrt{\left[s_{disk}^2+s_{cor}^2(\rho) \right] \ln\left[\frac{H_{res} K_{res}(\rho)I_{pB}(\rho)}{H_{pB}K_{pB}(\rho)I_{res}(\rho)} \right]} \label{eq:v0explicit}
\end{equation}
where $s_{disk} = \sigma_{disk}\, c/\lambda_0$ and $s_{cor} = \sigma_{cor}\, c/\lambda_0$. The above expression can be used
to measure the POS radial velocity of plasma pixel-by-pixel from the ratio between the $I_{res}$ and $I_{pB}$ intensity images, 
but only for regions where
\begin{equation}
	\frac{I_{pB}}{H_{pB}K_{pB}} \geq \frac{I_{res}}{H_{res}K_{res}}
\end{equation}
that, by looking at Eq. \ref{eq: DDratios}, simply corresponds to the condition that $D[V_0(\rho)] \leq 1$ as expected.

The advantage of the derived explicit expression (Eq. \ref{eq:v0explicit}) for the outflow speed $V_0$ is 
	also that this can be differentiated to estimate the dependence of the relative uncertainty $\Delta V_0/V_0$ on 
	the relative uncertainties on the other quantities. It turns out that
\begin{equation}
	\frac{\Delta V_0}{V_0} = \frac{s_{disk} \Delta s_{disk} + s_{cor} \Delta s_{cor}}{s_{disk}^2+s_{cor}^2} +
	\frac{\frac{\Delta I_0}{I_0} + \frac{\Delta R_H}{R_H} + \frac{\Delta I_{pB}}{I_{pB}} + \frac{\Delta I_{res}}{I_{res}}}{2\ln\left[\frac{H_{res} K_{res}I_{pB}}{H_{pB}K_{pB}I_{res}} \right]}.
\end{equation}
Considering that, by assuming for instance $\Delta s_{disk}/s_{disk} \simeq \Delta s_{cor}/s_{cor}$ the first therm 
	simply reduces to $\Delta s_{cor}/s_{cor}$, the above expression shows that the dependence of $\Delta V_0/V_0$ on the 
	relative uncertainties on the other quantities is not linear, because each one of these uncertainties is divided by a 
	factor that could be larger or smaller than unit depending on the considered heliocentric distance and coronal feature. The above 
	equation shows that the main uncertainties on $V_0$ are not only related with those on the widths of coronal ($\sigma_{cor}$) 
	and disk ($\sigma_{disk}$) line profiles, and on the uncertainties on the measured WL ($I_{pB}$) and UV ($I_{res}$) intensities,
	but also on the values of UV parameters $H_{res}$ and $K_{res}$, while the corresponding WL parameters $H_{pB}$ and 
	$K_{pB}$ are well know. As given in Eq. \ref{eq: uvbestapprox}, the former are related with the uncertainties on the 
	knowledge of the disk intensity $\Delta I_0 /I_0$ seen by the scattering atoms, and the Hydrogen ionization fraction 
	$\Delta R_H /R_H$, dependent on the electron temperature $T_e(\rho)$ \citep[see also][Fig. 11 and related discussion]{dolei2018}.

Some additional considerations about the described method are also important. First,
even if the estimate of the ratio between the $I_{res}$ and $I_{pB}$ intensities requires converting the WL intensity
from the usual relative units [$1/\bar{B}_{\odot}$] to the absolute units [phot cm$^{-2}$s$^{-1}$sr$^{-1}$], the constant
$H_{pB}$ in the above equations also contains the quantity $\bar{B}_{\odot}$, and in the end the measurement of $V_0$ turns out
to be independent on the value of $\bar{B}_{\odot}$. Second, all the above expressions assume to employ the
observed polarized brightness $I_{pB}$, but in principle also the total brightness $I_{tB}$ can be used, as far as
a good correction for the additional emission due to the F-corona is implemented. Third, it is also very important to
point out that at larger distances from the Sun (typically above $\sim 4$ \rsun\ in polar coronal holes and above
$\sim 10$ \rsun\ in coronal streamers) the \lya\ emission is dominated by the interplanetary emission, and a reliable
measurement of the outflow speed will be more difficult. Moreover, this background emission is not uniformly distributed
around the sky, and is also changing with the solar rotation \citep[see e.g.][Fig. 1]{bertaux2000} and with the solar 
activity cycle, as clearly shown by measurements acquired with the SOHO SWAN instrument \citep[see e.g.][Fig. 4]{quemerais2006}.

\section{Testing the quick inversion method} \label{sec: testing}

\subsection{Test with 1D radial profiles} \label{sec: 1Dtest}

The first test on the inversion method was performed by assuming from the literature analytic 1D radial profiles for 
different plasma parameters inside a typical coronal streamer and coronal hole at the minimum of solar activity
cycle. In particular, for this work the $n_e$ profiles were assumed from \citet{gibson1999} and \citet{cranmer1999}
respectively for coronal streamer and coronal hole, the $T_e$ profiles from \citet{vasquez2003} for both coronal streamer
and coronal hole, the $V_0$ profile from \citet{cranmer1999} for coronal hole, the $\xi$ profile from \citet{cranmer2020}
for coronal hole (no turbulent velocity was assumed for coronal streamer). For the outflow speed in coronal streamer
the following analytic expression was used
\begin{equation}
	V_0 (\rho) = \frac{74.3\,\rho -113.4}{0.135\,\rho+2.61}\,\,\,\mathrm{[km/s]}
\end{equation}
(with $\rho$ expressed in \rsun) that was derived from a fitting of measurements given by \citet{strachan2002} and 
\citet{noci2007}, and is valid only for $\rho > 1.53$ \rsun. Moreover, it assumed not only temperature isotropy,
but also thermodynamic equilibrium so that $T_e = T_k$. All these radial profiles are shown in Fig. \ref{Fig3}. 
Other constant quantities that have been assumed are the \lya\ disk intensity ($I_0 = 5 \times 10^{15}$ phot 
cm$^{-2}$s$^{-1}$sr$^{-1}$), and the $1/e$ half-width of the \lya\ chromospheric profile ($\sigma_{disk} = 0.34$ \AA).
\begin{figure}[t!]
	\centering
	\includegraphics[width=0.5\textwidth]{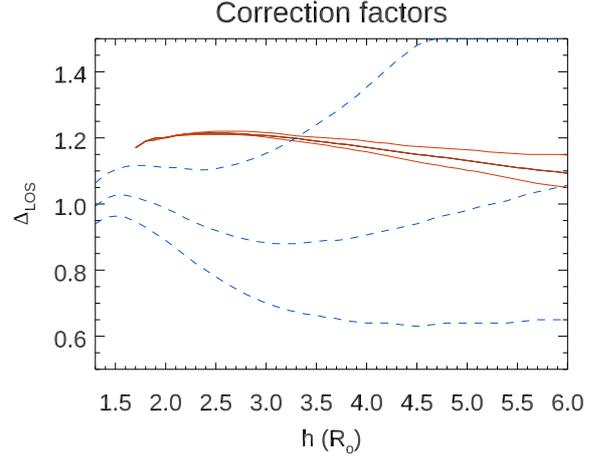}
	\caption{
		Radial profiles of correction factors $\Delta_{LOS}$ as derived for coronal streamer (solid red lines) and
		coronal hole (dashed blue lines). For each profile the upper/lower curves show the values of $\Delta_{LOS}$
		parameter resulting by introducing a $\pm5$\% error on the measured velocities.  
	}
	\label{Fig4}
\end{figure}

Starting from these input quantities, both the $I_{res}$ \lya\ and $I_{pB}$ intensities have been synthesized
at different altitudes $\rho$ and over $\pm 10$ \rsun\ along the LOS at each altitude, and then integrated along
the LOS; the resulting output radial profiles of $I_{res}$ \lya\ are in very good agreement with measurements
acquired by the SOHO UVCS instrument \citep{giordano2011}, as shown in Fig. \ref{Fig3} (middle left panel). 
In the integration along the LOS the analytic radial profiles for different plasma parameters have been used 
to derive at each altitude their LOS distribution, hence compute the $I_{res}$ emissivity and $I_{pB}$ intensity
emitted along the LOS for the integration. In particular, the electron temperatures have been used to determine
the fraction of neutral H atoms based on the ionization equilibrium curve provided by the CHIANTI spectral code 
\citep{dere2019}. Then, the $I_{res}$ \lya\ and $I_{pB}$ intensities  (Fig. \ref{Fig3}, middle panels) have 
been analysed by using the "quick inversion" technique described in the previous Section, to measure the outflow 
speed (Eq. \ref{eq:v0explicit}) and simulate the inversion of real data. Finally, a direct comparison between the 
input and the output outflow speed profiles allows testing and quantifying the accuracy of the inversion 
	method. Nevertheless, the use of approximate \lya\ and $pB$ expressions provided respectively in Eq. \ref{eq: uvbestapprox}
	and Eq. \ref{eq: wlbestapprox} inevitably introduces errors in the determination of the outflow speed $V_0$.
	Fortunately, because both approximate expressions are expected to overestimate the resulting intensity in the
	two spectral ranges, but the "quick inversion" method estimates the outflow speed from the ratio between 
	the two (Eq. \ref{eq:v0explicit}), these two errors tend to cancel out.

Hence, to optimize the velocity measurements we introduce here a correction factor $\Delta_{LOS}$, multiplying 
	in Eq. \ref{eq:v0explicit} the ratio between $I_{pB}$ and $I_{res}$ values by this factor. Then, by iterating over 
	$\Delta_{LOS}$ values in the range between $0.5-2.0$, we measured at any altitude the value of this correction factor 
	making the outflow velocities measured with the "quick inversion" method coincident with the assumed input values, 
	both for coronal streamer and coronal hole cases. This allowed us to estimate at any altitude the needed correction 
	for the ratio between $I_{pB}$ and $I_{res}$. Values of $\Delta_{LOS}$ derived with this analysis are shown in 
	Fig. \ref{Fig4}, together with the corrections needed by assuming an uncertainty by $\pm 5$\% on the outflow speed.
	Results show first of all that the uncertainties in coronal holes (dashed blue lines) are expected to be smaller with 
	respect to coronal streamers (solid red lines), because a larger interval of possible $\Delta_{LOS}$ values allow 
	measuring in output the same input velocities with an uncertainty below 5\%. Moreover, a constant value $\Delta_{LOS} \simeq 1.2$
	provides the best compromise both for coronal streamers at lower altitudes and coronal holes at higher altitudes. Hence,
	because our purpose is to provide a general method applicable to any coronagraphic image and any coronal structure, in what 
	follows we will assume this value as the best compromise.

Finally, the outflow velocities $V_0$ resulting by applying Eq. \ref{eq:v0explicit} with a constant correction 
	factor $\Delta_{LOS} = 1.2$ are given in the bottom right panel of Fig. \ref{Fig3} (dotted red and blue lines) and 
	compared with input velocities for a coronal streamer (solid red line) and a coronal hole (dashed blue line)
Results from this simple 1D analysis show that in general the speeds measured with the "quick inversion method"
will reproduce the input velocity profiles with very small errors (less than $\sim 20$ \kms), both in coronal 
streamers and coronal holes. More in details, larger discrepancies could result only in the inner regions of 
coronal streamers below $\sim 2$ \rsun, where the resulting speeds could be overestimated. The possible reason 
for these discrepancies will be discussed later on.

\subsection{Test with 3D MHD simulations} \label{sec: 3Dtest}

In order to test also the effects of LOS integration once the hypothesis of cylindrical symmetry is removed,
in this work a second test was performed based on 3D numerical MHD simulations. In particular, the 3D datacubes
with all the main plasma physical parameters in the inner corona are freely distributed by the Predictive
Science Group\footnote{\href{http://www.predsci.com/hmi/home.php}{See Predictive Science Inc. webpage.}}. 
These reconstructions start from the 
photospheric magnetic field measurements acquired by the HMI instrument on-board the SDO mission 
\citep{scherrer2012} and are based on the well-established Magnetohydrodynamic Algorithm outside a Sphere 
(MAS) model \citep[see e.g.][]{mikic1999, linker1999}. The 3D datacubes were selected, downloaded, and
managed by using the FORWARD data package freely distributed with SolarSoftware \citep{gibson2016},
that also allows to create synthetic images in many different wavebands simulating the view from the
Sun-Earth line for a specific date. This datapackage, thanks to a collaboration during a dedicated
"ISSI International Team", was upgraded in order to include also the computation of the \lya\ coronal
emission, based on the method originally developed by \citet{fineschi1993} to measure the coronal
magnetic fields by taking advantage of the modification induced in the linear polarization of this
emission line by the Hanle effect \citep{bommier1982}. In particular, the synthetic \lya\ intensity
is calculated with the same Equations given in Sections 2.1 and 2.2 of \citet{khan2011}, by performing
the full integration over the solid angle of the solar disk and along the LOS, and by approximating
the wavelength integration with the same expression given here by the convolution of two Gaussian 
profiles (Eq. \ref{eq:convol3}).

\begin{figure*}[t!]
	\centering
	\includegraphics[width=0.8\textwidth]{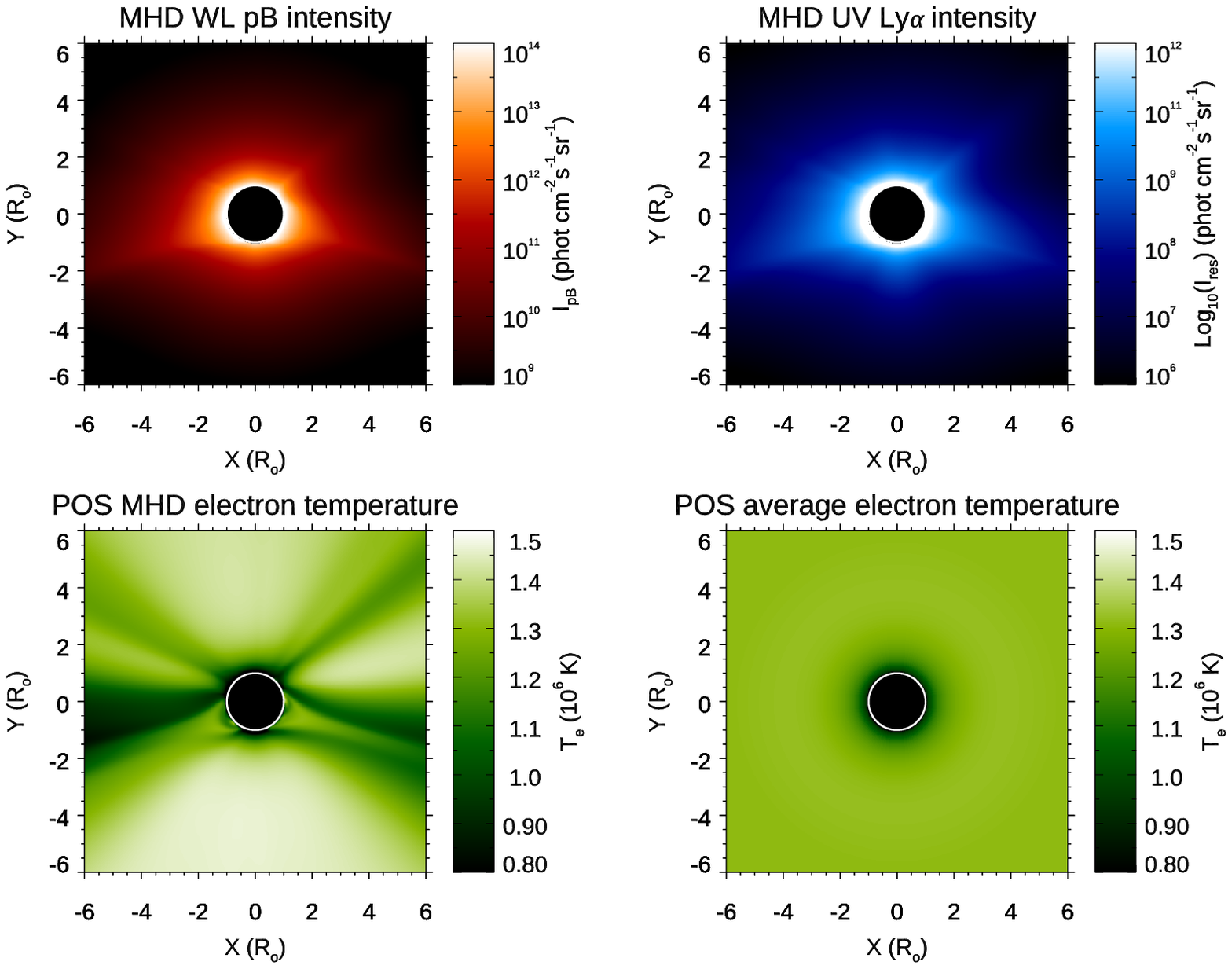}
	\includegraphics[width=0.8\textwidth]{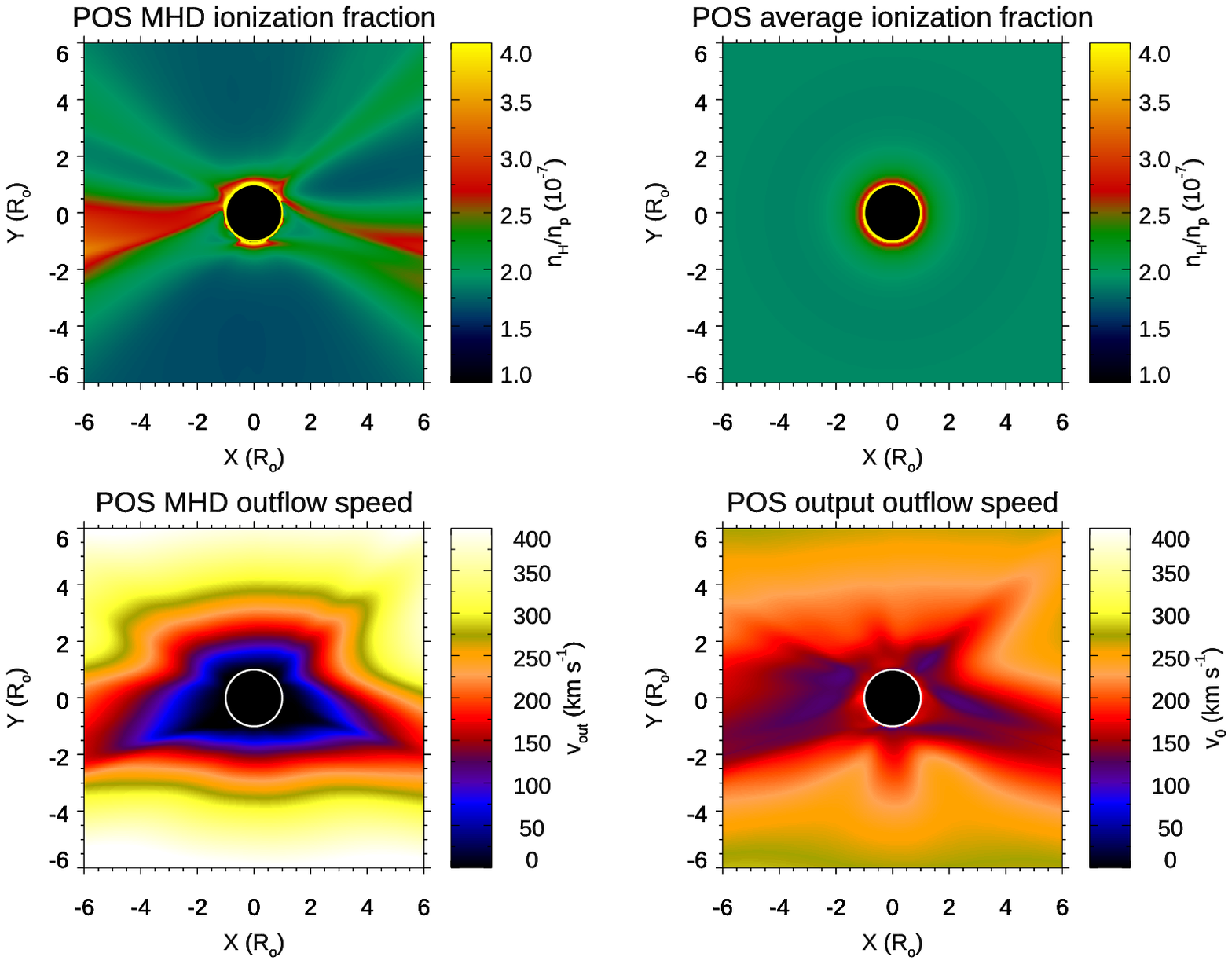}
	\caption{
		From top to bottom, and from left to right: synthetic WL (pB) and 
		UV (\lya) images, the POS temperature distribution and the resulting
		averaged profile, the corresponding POS distribution of neutral H
		atoms and the resulting averaged distribution, and input and output
		outflow speeds on the POS. These panels refer to the simulation performed
		close to the minimum of solar activity cycle.
	}
	\label{Fig5}
\end{figure*}
For this work, we selected two specific dates corresponding to the times when two
different Total Solar Eclipses occurred on Earth (in order to have also possible comparisons on 
the appearance of the real inner solar corona in the WL images acquired from the ground). In
particular, the selected dates correspond to the eclipses that occurred on 2017 August 21 (17:33 UT)
near the minimum of solar activity cycle, and on 2012 November 13 (22:13 UT) near the maximum
of solar activity cycle. This allows to test the inversion method in two different conditions,
when the 3D structure of the solar corona is closer (near minimum) or farther (near maximum) from the cylindrical symmetry. 
For each one of these two dates, the synthetic WL ($pB$) and UV (\lya) images were built taking into
account the LOS integration. Then, in order to perform the data analysis, and to take into account
that the inversion of the real data will be performed without a clear knowledge of different 
temperatures in the real corona, the plasma temperatures in the model were extracted on the POS 
and averaged over all latitudes, in order to employ only the same radial temperature profile at 
all latitudes. This 2D average temperature maps were thus used to reconstruct (based on the
usual assumption of ionization equilibrium) the 2D distribution of neutral H atoms on the POS.
Finally, the "quick inversion method" was applied pixel-by-pixel to the synthetic images to
determine the 2D distribution of outflow speed on the POS, and the results were compared with
the real plasma velocities extracted from the model on the POS. For this 3D test a
correction factor $\Delta_{LOS} = 1.2$ as derived above was used in the analysis.

\begin{figure*}[t!]
	\centering
	\includegraphics[width=0.8\textwidth]{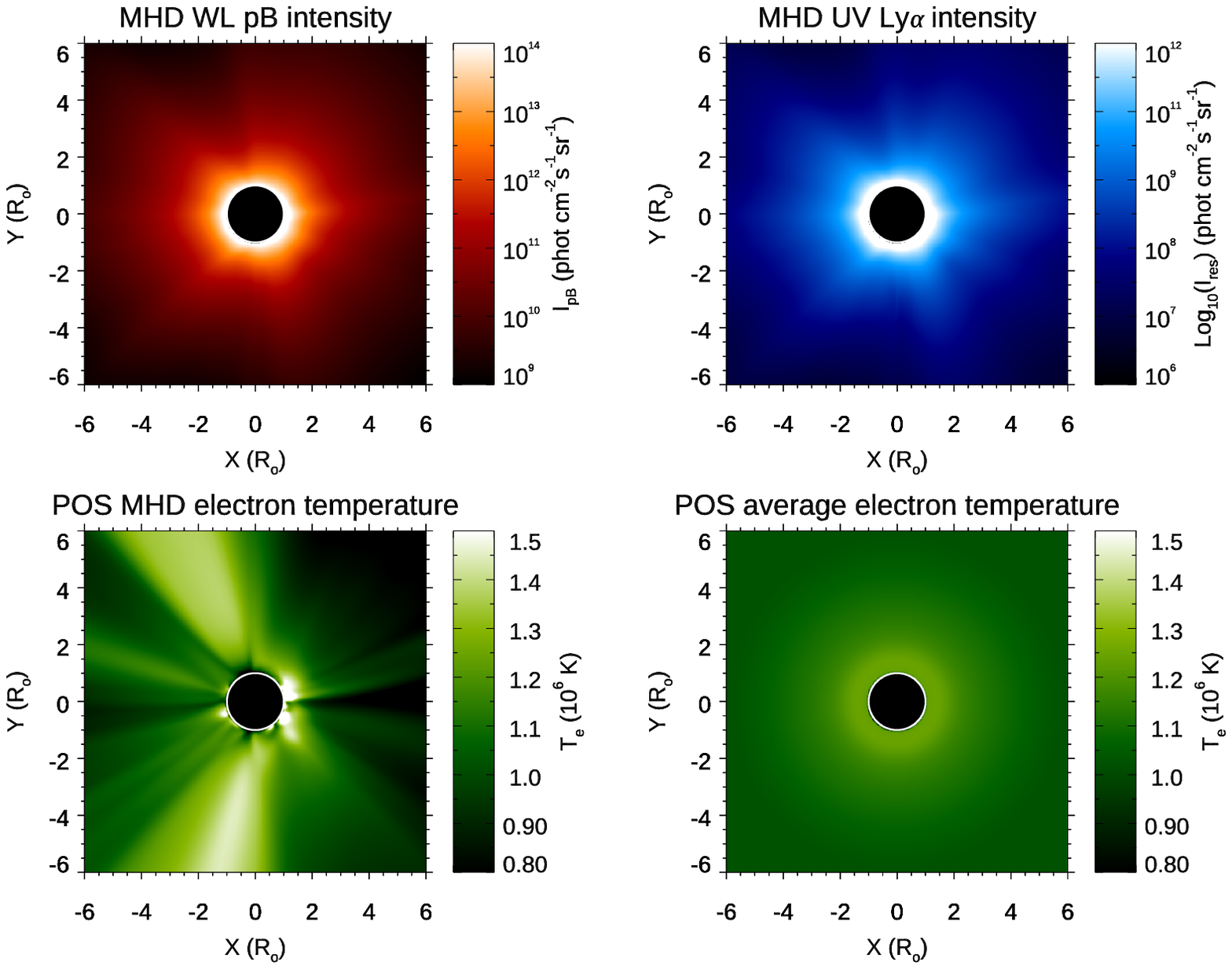}
	\includegraphics[width=0.8\textwidth]{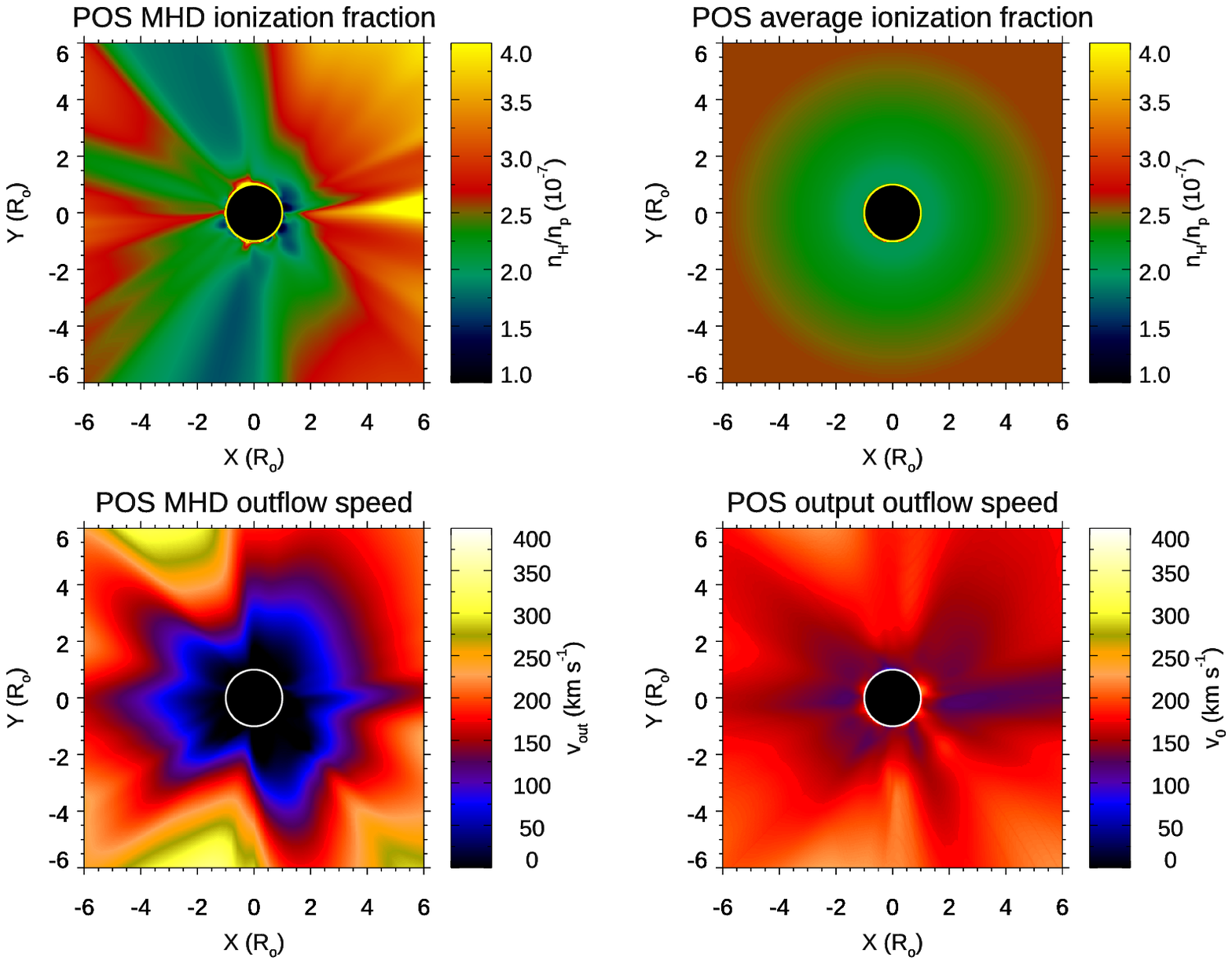}
	\caption{
		Same as for Fig. \ref{Fig5}, regarding the simulation performed close
		to the maximum of solar activity cycle.
	}
	\label{Fig6}
\end{figure*}

Results are shown in Fig. \ref{Fig5} and Fig. \ref{Fig6} for the solar minimum and solar maximum 
conditions, respectively. The UV \lya\ intensities were computed by assuming a chromospheric
intensity $I_0 = 5.24\cdot 10^{15}$ phot cm$^{-2}$s$^{-1}$sr$^{-1}$. The same $I_0$ intensity was assumed 
here for both cases independently on the phase of the solar activity cycle, because by using exactly the 
same value also for the data inversion, this makes the results independent on the value of $I_0$.
Bottom panels of Fig. \ref{Fig5} and Fig. \ref{Fig6} show a direct comparison between the 2D distribution of 
POS outflow speed in the MHD model (left) and the output POS speed as derived with the "quick inversion method".
Obviously, the plasma physical parameters in the MHD numerical models have for each pixel in 
the 2D synthetic UV and WL images a 3D distribution along the LOS which is in general unknown.
In particular, the MHD model has different LOS velocities that are not shown in the bottom left
panels of Fig. \ref{Fig5} and Fig. \ref{Fig6}. The inversion method is expected to be affected 
by this LOS integration effects as much as the real 3D distribution of coronal plasma parameters 
departs from a cylindrical symmetric distribution.

In general, Fig. \ref{Fig5} and Fig. \ref{Fig6} show that the output maps of outflow speeds allows
to identify the 2D distribution of coronal regions characterized by fast and slow wind streams.
This is more evident for the solar minimum condition (Fig. \ref{Fig5}), where the clear fast/slow 
wind dichotomy is well reproduced around polar/equatorial regions, respectively. The
situation becomes more complex for the solar maximum condition (Fig. \ref{Fig6}), where in any
case the morphological distribution of different wind streams is reproduced. On the other hand, 
the main problems are related with the absolute values resulting from the analysis. The measured
velocities appear to be slightly underestimated in the outer corona, and also overestimated in 
the inner corona typically below $\sim 2$ \rsun\ (see also reference values in the bottom right 
panel of Fig. \ref{Fig3}). A more quantitative comparison is shown by
different panels of Fig. \ref{Fig7}, showing that at solar minimum (left column) the velocities
are slightly underestimated at 5.0 \rsun\ (bottom left panel) and overestimated at 2.5 \rsun\
(top left panel), while at solar maximum (right column) the velocities are in quite good agreement 
at 5.0 \rsun\ (bottom right panel) and significantly overestimated at 2.5 \rsun\ (top right panel).
The possible reason for these discrepancies will be discussed later on.

\subsection{Test with real observations} \label{sec: realdata}

A further and last test of the "quick inversion method" described here was performed based on real 
data acquired by the LASCO \citep{brueckner1995} and UVCS \citep{kohl1995} instruments on-board
SOHO. Again, two different time periods were selected, to test the data analysis during two
different phases of the solar activity cycle. In particular, the daily LASCO-C2 $pB$ images
were downloaded according to the most recent version of instrument radiometric calibration
\citep[see e.g.][]{lamy2020} to avoid possible residual stray light in LASCO $pB$ images 
that could be responsible for an overestimate of $I_{pB}$, and finally for an overestimate of the
outflow speed according to Eq. \ref{eq:v0explicit}. For the test described here we focused
over the time periods corresponding to Carrington Rotations 1910 (from
1996 June 1, 11:58 UT to June 28, 16:43 UT, hence at the minimum of solar
activity cycle), and 1963 (from 2000 May 17, 02:55 UT to June 13, 07:51 UT,
hence at the maximum of solar activity cycle). The resulting average WL $pB$ images are
shown in the left panels of Fig. \ref{Fig8}.

\begin{figure*}[t!]
	\centering
	\includegraphics[width=0.75\textwidth]{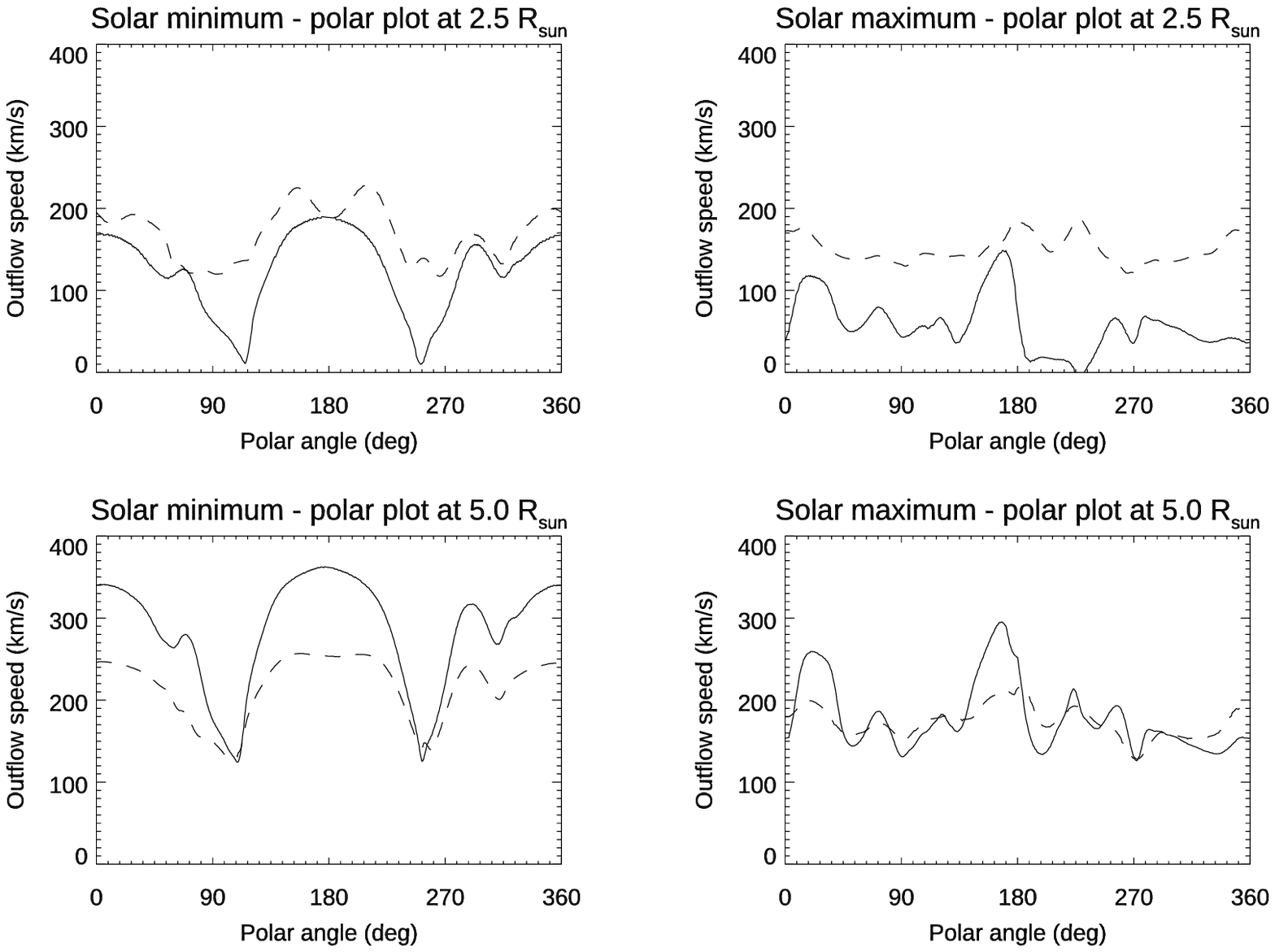}
	\caption{
		Polar plots of the POS outflow speeds as provided in input from the model (solid lines)
		and as resulting in output from the "quick inversion method" (dashed lines), for the 
		solar minimum (left column) and solar maximum (right column) conditions. The plots show
		the polar distributions at constant heliocentric distances of 2.5 \rsun\ (top row) and
		5.0 \rsun\ (bottom row).
	}
	\label{Fig7}
\end{figure*}

\begin{figure*}[t!]
	\centering
	\includegraphics[width=\textwidth]{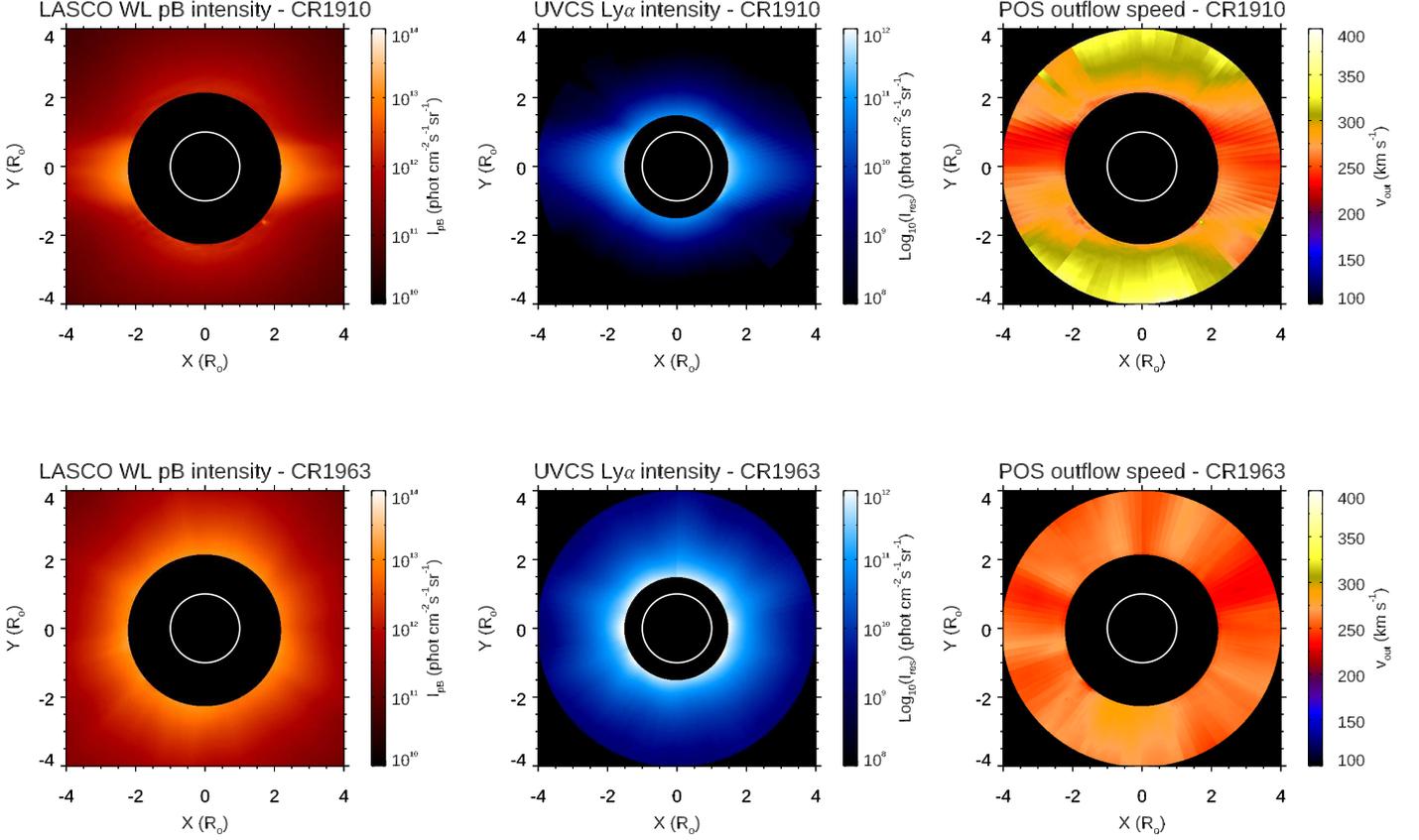}
	\caption{
		The average LASCO WL $pB$ (left panels) and UV \lya\ (middle panels) images obtained for Carrington Rotations 1910
		(top row) and 1963 (bottom row), and the corresponding distribution of POS speeds as derived with the "quick
		inversion method" (right panels).
	}
	\label{Fig8}
\end{figure*}

For the same time intervals, the averaged \lya\ coronal intensity maps were built by collecting all 
together the spectroscopic observations acquired by the UVCS instrument at different latitudes and altitudes, 
and then by interpolating and extrapolating the \lya\ intensities with power law fitting to fill the data gaps; 
a similar method was also applied by \citet{bemporad2017} to analyze one full solar rotation period with the 
UVCS \lya\ acquired with the so-called synoptic observations \citep[see e.g.][]{giordano2008}. 
In the analysis presented here the \lya\ intensities were extrapolated in the range between 1.5 and 4.0 \rsun;
resulting average UV \lya\ images are shown in the middle panels of Fig. \ref{Fig8}. Because on the other hand 
the LASCO $pB$ images are going from 2.1 to 6.0 \rsun, the combined analysis with the "quick inversion method"
provides solar wind velocity maps in the range between 2.1 and 4.0 \rsun, as shown in the right panels of 
Fig. \ref{Fig7}.

In particular, the "quick inversion method" has been applied here by assuming chromospheric \lya\ intensities of 
$I_0 = 5.66\cdot 10^{15}$ and $I_0 = 8.24\cdot 10^{15}$ phot cm$^{-2}$s$^{-1}$sr$^{-1}$ as measured by the 
SOLSTICE satellite \citep{rottman2001} during CR1910 and CR1963, respectively, and in agreement with most
recent data re-calibration \citep{machol2019}. Coronal electron temperature radial profile has been assumed at 
all latitudes equal to an average between the profiles given by \citet{cranmer1999} and \citet{gibson1999} for 
the polar and equatorial regions, respectively. The resulting 2D electron temperature map has 
been converted into a neutral Hydrogen map by assuming again ionization equilibrium from CHIANTI. An
interplanetary \lya\ intensity by $I_{interp} = 3.0\cdot 10^{7}$ phot cm$^{-2}$s$^{-1}$sr$^{-1}$ 
as provided by \citet{kohl1997} has been subtracted from both the reconstructed coronal \lya\ images.
Again, the results for the outflow speed have been optimized by assuming a correction factor 
$\Delta_{LOS} = 1.2$ as previously derived.

The outflow speed maps from application of the "quick inversion method" are shown in the right panels of Fig. \ref{Fig8} 
for Carrington Rotations 1910 (top) and 1963 (bottom). The plots show that in both cases it was possible to derive
the 2D distribution of POS outflow velocities, and the resulting maps shows very well the locations 
of higher and lower velocities in nearby solar wind streams. In particular, while the outflow speed map
for the minimum of solar activity cycle (top right panel) clearly shows the classical bi-modality of solar wind 
(with fast/slow wind streams limited at the polar/equatorial regions), the map around the maximum of solar
activity cycle (bottom right panel) shows faster/slower streams located at all latitudes, as expected.
Moreover, the very high velocities observed during solar minimum around the polar regions are never reached 
at any latitude around solar maximum: this result is also confirmed by most recent solar wind speed measurements 
provided in different phases of the solar activity cycle and obtained in the extended corona with Fourier
filtering applied to LASCO-C3 images \citep{cho2018}, and much farther from the Sun with radio scintillation
measurements \citep{sokol2013}. In any case, we expect that these measurements will be partially affected by 
errors similar to those that have been found from the analysis of synthetic images and described before 
(end of Section \ref{sec: 3Dtest}). In the near future we plan to apply this method to re-analyze 
all the UVCS observations and provide to the community a catalogue of 2D solar wind speed maps for many 
different Carrington rotations between 1996 and 2002.

\section{Discussion \& conclusions} \label{sec: summary}

In this work the possible application of an inversion technique was described and tested with analytic profiles, 
numerical MHD simulations, and real observations. It is also important to notice that the direct ratio technique 
(called "quick inversion method") not only will provide with a few steps reliable determination of the plasma 
outflow speeds in the corona, but will also have many advantages with respect to the classical "full inversion 
method", that are listed here.
\begin{itemize}
\item Possible uncertainties related with the derivation of coronal electron densities are entirely avoided, 
because the velocities are derived directly from the intensity ratio in the two spectral bands.

\item Possible smaller scale inhomogeneities in the outflow speed in the radial direction are detectable down 
to the projected pixel size, because the velocity in each pixel is independent of the 
nearby pixels, while the full LOS integration method is based on power law density profiles derived with 
geometrical assumptions, and a similar assumption is made also in the outflow speed profiles, thus 
significantly smoothing the possible pixel-by-pixel speed inhomogeneities.

\item Images acquired during the transit of small (e.g. blobs, jets, etc.) or large (Coronal Mass Ejections, 
shocks, etc.) scale impulsive and transient events can be analysed as well with the
direct ratio technique, because the method derives a measure of the outflow speed pixel by pixel independently, 
which is not true for the full inversion method which is based on power law fitting of the density profile 
and assumptions about the LOS integration that are not applicable to localized plasma features. 

\item The projected radial extension of the instrument field of view in coronagraphic images is used 
entirely from the inner to the outer edges of the images, while the method performing the full integration 
along the LOS requires assuming for a given thickness along the LOS the velocities measured on the POS, 
and this assumption cannot be made for pixels located closer to the outer edge of the images. 

\item Possible residual instrumental artefacts not removed by the absolute radiometric 
calibrations tend to cancel out in the direct intensity ratio, as far as the pattern of these artefacts 
(e.g. stray light, residual vignetting, etc.) are similar in the two channels; this is something that 
one may expect for the Metis instrument \citep{liberatore2021}, considering that a significant amount of the 
optical path is shared for the two channels, while the same is not necessarily true for other instruments.
\end{itemize}
Nevertheless, the analysis described here also shows that the "quick inversion method" has also significant
limitations. In particular, even if the method is able to reproduce the 2D distribution of higher and lower
velocities on the POS, the absolute values are expected to be slightly underestimated in the outer corona
(above $\sim 2$ \rsun) and significantly overestimated in the inner corona (below $\sim 2$ \rsun). The
sources of these two different uncertainties are totally different and are briefly discussed here. In the outer corona, 
as the solar wind velocities increasing with altitude approaches the value of about $\sim 300$ \kms, the
Doppler dimming technique with \lya\ line starts to be insensitive to higher velocities. The reason is that
above this velocity the values of the Doppler dimming coefficients (Fig. \ref{Fig1}) asymptotically goes to
zero, making the measurement more and more uncertain. This is particularly true in
coronal holes (see Doppler dimming values in the bottom left panel of Fig. \ref{Fig3}), but also velocities
measured in coronal streamers are affected at higher altitudes (bottom left panel of Fig. \ref{Fig7}). It
is important to notice that this limit will affect the wind speed measurements obtained both with
the "quick inversion" and the "full inversion" methods, because this limitation is intrinsically 
related with the variations of Doppler dimming coefficient as a function of the outflow speed. 

On the other hand, the uncertainties in the velocity determinations in the inner corona are mainly 
due to the approximations performed in the "quick inversion method", and related to
the LOS integration that is neglected by the method. In particular, for the "quick inversion" it was 
assumed that the integration over the solid angle subtending the solar disk can be simply factorized as 
expressed by eq. \ref{eq:omega}, usually referred as "point source" approximation. Obviously, this
approximation fails for regions in the inner corona, where different values of this solid angle in the
integration along the LOS need to be taken into account. This source of error is peculiar of the "quick
inversion" and is not present in the "full inversion" method. Hence, this leads us to conclude that
while the method is applicable for instance to Metis data (considering that the instrument FOV will
never observe the inner corona below 1.7 \rsun), care must be taken by applying this method also to
future images that will be acquired by the LST instrument, whose FOV will extend down to the solar limb.
A possible modification of the "quick inversion method" to remove this source of errors will be
considered in a future work. Please also notice that this effect likely led to an overestimate of
the solar wind velocities in the inner corona as published by \citet{bemporad2017}. 
Despite these source of uncertainties, the "quick method" described here could be used 
	in principle also to discriminate from real data analysis between different models for the solar
	wind acceleration. To demonstrate this capability will need to create synthetic data starting from 
	different numerical models (with different physical treatments of the solar wind), and then to invert 
	these data with the method described here to compare finally different results. This interesting 
	analysis goes beyond the purposes of the present work, and will be considered as a future development.

Before concluding, it is also interesting to point out that, when large or small scale parcels of plasma propagating 
through the corona (e.g. blobs, jets, CMEs, etc.) will be detected, the explicit expression for the outflow speed 
(Eq. \ref{eq:v0explicit}) provided here can be reversed to measure the evolution of plasma temperatures. In fact, 
if we assume that by tracking the plasma feature (propagating inward or outward the corona) in coronagraphic
images it is possible to measure (on the POS) the radial velocity profile $V_0(\rho)$ as a function of distance
$\rho$ (or as a function of time), this also provides the evolution of the H ionization fraction which is 
given by
\begin{equation}
	R_H[T_e(\rho)] = \frac{H_{res}}{H_{pB}}\frac{I_{res}(\rho)}{I_{pB}(\rho)}\frac{K_{pB}(\rho)}{h(\rho)}
	\frac{\sqrt{\sigma_{disk}^2+\sigma_{cor}^2(\rho)}}{\exp\left[ -\frac{V_0^2(\rho)}{(\sigma_{disk}^2+\sigma_{cor}^2(\rho))c^2/\lambda_0^2}\right]}.
\end{equation}
The application of the above expression requires first of all to measure the excess brightnesses of moving
plasma features both in WL and UV (in order to remove contamination from the emitting plasma aligned with
the external corona along the LOS); moreover, it is also necessary to make some assumptions on the evolution
of quantity $\sigma_{cor}(\rho) = \lambda_0/c \sqrt{2 k_B T_k(\rho)/m + \xi^2(\rho)}$. Once with the above 
expression the $R_H[T_e(\rho)]$ curve is measured, this can be reversed to measure the $T_e(\rho)$ curve,
considering that for $T_e$ between $10^6$ and $10^8$K the ionization equilibrium curve provided by the 
CHIANTI spectral code \citep{dere2019} can be fit to about 10\% accuracy by
\begin{equation}
	T_e \simeq 0.59 \cdot 10^6 \,\,R_H^{-0.9407}
\end{equation}
as recently provided by \citet{cranmer2020} (Eq. 7). Hence, the "quick inversion method" described here
has the advantage to be applicable in theory also to investigate the thermodynamic evolution of plasma
erupting from the Sun at any spatial scale from large- to small-scale eruptions. The method can be applied 
under the hypotheses that 1) ionization equilibrium is still present, and 2) that the observed \lya\ emission 
is entirely due to radiative excitation alone. These two hypotheses are not necessarily verified in the whole 
volume of CMEs \citep[see discussions by][]{susino2018, bemporad2018, pagano2020}, but could be verified for
small-scale less energetic phenomena such as propagating plasma blobs or density inhomogeneities. This possible 
application of the "quick inversion method" will be tested with MHD numerical simulations and real observations 
in future works.

\begin{acknowledgements} 
	The authors thank S. Fineschi and R. Susino for providing the upgraded version of the FORWARD
	data package simulating the \lya\ coronal emission. F. Frassati is supported through the Metis 
	programme funded by the Italian Space Agency (ASI) under the contracts to the co-financing National 
	Institute of Astrophysics (INAF): Accordo ASI-INAF n. 2018-30-HH.0
\end{acknowledgements}

\bibliographystyle{aa}
\bibliography{biblio}

\end{document}